# Josephson Effect in Metal - Polyimide - Metal Sandwich Structures

A.N. Ionov, V.A. Zakrevskii

A.F. Ioffe Physico-Technical Institute, RAS , 194021, St. Petersburg; Russia

J.M. Lazebnik

B.P. Konstantinov Nuclear Institute, RAS, 18835, Gatchina, Russia

R. Rentzsch

Institut für Experimentalphysik, Freie Univesität Berlin, D-14195 - Berlin, Germany

V.M. Svetlichny

Institute of Macromolecular Compounds, RAS, 199004, St. Petersburg; Russia

We have observed that films of amorphous polyimide with a thickness $d \leq 1.8$ μm placed between two metallic electrodes become highly conducting with no applied electric field, and at a low pressure which was necessary only to achieve reliable electrical contact with the metallic electrodes. We observed for the first time a Josephson effect in a sandwich of a superconducting metal - polyimide film - superconducting metal at $T \leq T_c$ of the superconducting electrodes where $d/\xi_o >> 1$ (here $\xi_o$ is the coherence length of the superconducting electrodes and $d$ the thickness of the polyimide film).

It is well known that the Josephson effect is the physical basis for many quantum devices [1]. The high sensitivity to changes of the magnetic field in SQUIDs allows one to use them in a wide variety of different applications from physics and technology, to medicine and biomagnetism.

In this letter we demonstrate a new construction based on a conductive film of amorphous polyimide placed between two superconducting electrodes which exhibits the Josephson effect.

In [2] it was shown that films of polyimide with a thickness $d$ = 2 - 3 μm placed between two metallic electrodes become highly conducting in a relatively small electric field ($E < 10^3$ V/cm), if a small ($P < 10^5$ Pa) uniaxial mechanical pressure is applied to the metallic electrodes. It was also shown that if the metallic electrodes were in the superconducting state, an effective resistance of zero was recorded.

In our experiments we used polyimide poly-[4,4´-bis-(4-N-phenoxy)diphenylsulfone] imide 1,3-bis-(3,4-dicarboxyphenoxy)benzene (R-SOD) obtained by using a single-stage high temperature imidization in solution in N-methylpyrrolidone (MP) [3]. The polyimide films, with a thickness of 0.7 – 3.0 μm, were prepared by a centrifugal method by deposition of a drop of 20 % wt. solution of polyimide (R-SOD) in MP on a polished metallic bulk electrode of 10 mm diameter. After reaching the intended film thickness, the film was heated in air for 2 h at a constant temperature of 420 K to remove the polymer solvent. The softening temperature of the polyimide was about 450 K, and its elastic modulus was around 2400 MPa at 300 K. The second electrode, 5 mm in diameter, was an electrical contact pressed on top of the polymer. The electrodes had a mean surface roughness less than 80 nm and a purity of better than 99.99 %. To exclude tilting between the two electrodes, the top electrode, was surrounded by an insulating ring with an external diameter of 10 mm so that the top electrode and ring had the same surface i.e. without any protuberances. In some experiments we placed a dielectric ring of 1μm thickness and with external diameter of 5 mm between the two contacts. The ring had an internal diameter of 2 mm which was filled with polyimide. The formation of the conducting state in polyimide was tested in the same way as described in [4]. We observed that the high conductivity of sandwich structures with a polyimide film of $d \leq$ 1.8 μm arises in zero applied electrical field and at the lowest pressure ($P < 10^5$ Pa) which was necessary to achieve a reliable electrical contact with the metallic electrodes. High conductivity was found in a broad temperature range from 4.2 to 300 K. The same behavior was recently observed for poly(3,3´´-phthalidylidene-4,4´´- biphenylylene) films (PPB) [4] which also belong to the aromatic class of polymers such as polyimides. The highly conducting state in polymers presupposes that there is a high concentration of charge carriers with high mobility. Electrification of polymers in contact with metals, i.e. charge transport across the interface, is a well-known phenomenon studied by various authors [5]. The charge acquired by a polymer from a metal that touches it depends on the type of polymer, on the type of the metal and on the contact area.

The total electrically active area of the contact can be less than its apparent area. This is because surfaces always have roughness and when touched together, they make electrical contact only at some spots. Under the action of an external mechanical pressure the contact area is changed. In our sandwich structure we observed a correspondence between the contact area and the total value of the resistance. The smaller the contact area, the higher the resistance of the sandwich structure. Two - terminal resistance measurements with constant current ($I_{max} < 100$ μA) showed that the resistance of the sandwich structure at room temperature usually was in the range of 0.05 Ω for a “large” contact area and up to 10 Ω for a “small” contact area. As the temperature decreased, slow fluctuations of the resistance were observed down to the temperature $T_c$, where the electrodes (Sn or Nb) undergo a transition into the superconducting state. At $T \leq T_c$ the resistance fluctuations disappear sharply.

We now discuss the temperature dependence of the sandwich structures with low resistance. Figures 1a and 1b show typical temperature dependencies of resistance for a superconductor – polyimide - superconductor sandwich (SPS) with a film thickness $d$ = 1.5 μm. We used bulk electrodes of Sn and Nb which are superconducting at $T_c$ = 3.72 and 9.2 K, respectively. The two - terminal resistance measurements were done at a constant current with $I \leq 100$ μA. The sharp drop in the resistance at $T_c$ shown in the figures is similar to a superconducting transition of the SPS structure with the following peculiarities:

i) at small enough current the electrical resistance at $T \leq T_c$ was unmeasurably low (Figures. 1a, and 1b);

ii) at larger currents, the measured resistance remained finite and depended on the absolute value of electrical current, especially at the lowest temperatures (Fig. 1a).

A typical current – voltage characteristic of a SPS structure is shown in Figure 2. It looks like that for a superconducting constriction formed by a weak link: e.g. a normal metal with proximity effect between two superconducting electrodes (SNS) or a point contact between two superconductors (ScS) [6]. We investigated the existence of a weak link between the superconducting electrodes due to the proximity effect in our polyimide film. To that end we measured the voltage drop over the SPS sandwich structure at constant electrical current just above $I_c$ by sweeping a low transverse magnetic field ($H \perp I$). If a weak link indeed exists, one should observe oscillations of the critical current $I_c$ as a function of the magnetic field as predicted by Josephson [1]. Alternatively, at constant current the amplitude of the voltage oscillations ΔU can be measured, which strongly depend on the ratio $I/I_c \geq 1$ [7] which is experimentally easier to measure as function of the magnetic field. In Figure 3 the voltage oscillations are shown as function of the external magnetic field. In this particular experiment we used an "unsymmetrical" sandwich of Sn-P-Nb. The principal results were the same as in a "symmetrical" sandwich, which consists of the same material for both electrodes. The shape and period of the voltage oscillations are similar to the "classical" Josephson effect [8,9]. In our system the coherence lengths were $\xi_o$ = 250 nm for Sn and $\xi_o$ = 38 nm for Nb electrodes [10]. We did not observe such voltage oscillations if one of the electrodes was in the normal state ($3.72 < T \leq 9.2$ K). Therefore, at first sight this seems to suggest that the superconductive state in the polyimide film is due to a proximity effect.

In this connection the following questions remain open:

i) what type of weak link is responsible for the Josephson effect;

ii) what type of proximity effect inside the polymer can occur over such long distances as $d = 1.5$ μm, where $d/\xi_o >> 1$ holds ($\xi_o$ is the coherence length of the superconducting electrodes).

Further investigation is needed to throw light on the nature of the highly conducting state in polyimide.

**In conclusion** we observed for the first time a Josephson effect in a sandwich of a superconducting metal – polyimide film - superconducting metal at $T \leq T_c$ of the superconducting electrodes. The thickness of the polyimide film in this case can be as high as 1.5 μm. We speculate that the Josephson effect is due to a new type of long range proximity effect in polymers.

**Acknowledgments**

We thank Prof. S. Ginsburg for discussion and Mrs. V. Sarigina for her help with the sample preparation.

## References

[1] B.D. Josephson, Phys. Lett. **1**, 251, (1962); Advances in Physics **14**, 419 (1965).

[2] A.M. El´yashevich, A.N. Ionov, V.V. Kudryavtsev, M.M. Rivkin, V.M. Svetlichnyi, I.E. Sklyar, and V.M. Tuchkevich, Vysokomolekulyarnye Soedineniya , **B35**, 50 (1993) [Polymer Science, **35**, 40 (1993)].

[3] V.M. Svetlichnyi, T.I. Zhukova, V.V. Kudriavtsev, G.N. Gubanova, V.E. Yudin, A.M. Leksovskii, Polymer Engineering and Science, **35**, 1321 (1995).

[4] V.A. Zakrevskii, A.N. Ionov, A.N. Lachinov, Pis'ma Zh. Tekh. Fiz., **24**, 89 (1998) [Tech. Phys. Lett., **24**, 539 (1998)].

[5] J. Lowel, A.C. Rose-Innes, Adv. Phys., **29**, 947 (1980) and references therein.

[6] A.C. Rose-Innes and E.H. Rhoderick, in *Introduction to Superconductivity* (Pergamon Press 1969) Chap. 8.

[7] M.N. Omar and R. De Bruyn Ouboter, Physica **32**, 2044 (1966).

[8] M.N. Omar et.al, Physica, **34**, 525 (1967).

[9] R. De Bruyn Ouboter, in *Progress in Low Temperature Physics*, edited by C.J. Gorter (North-Holland, Amsterdam, 1970), Chap. 6.

[10] see for instance in *Superconductivity*, edited by R.D. Parks (Marcel Dekker, NY, 1969), Vol. 1.

## Figure Captions

Figure 1a. The temperature dependence of resistance for a sandwich structure of Tin – Polyimide film – Tin at different values of the current.

Figure 1b. The temperature dependence of resistance for sandwich structure of Niobium – Polyimide film – Niobium.

Figure 2. Current vs. voltage for a Tin – Polyimide film – Tin sandwich at $T$ = 1.79 K.

Figure 3. The voltage oscillations $U$ vs. magnetic field $H$ for a sandwich structure Tin – Polyimide film – Niobium at $T$ = 2.8 K and at $I/I_c \geq 1$, ($I_c$ = 163 $\mu$A).

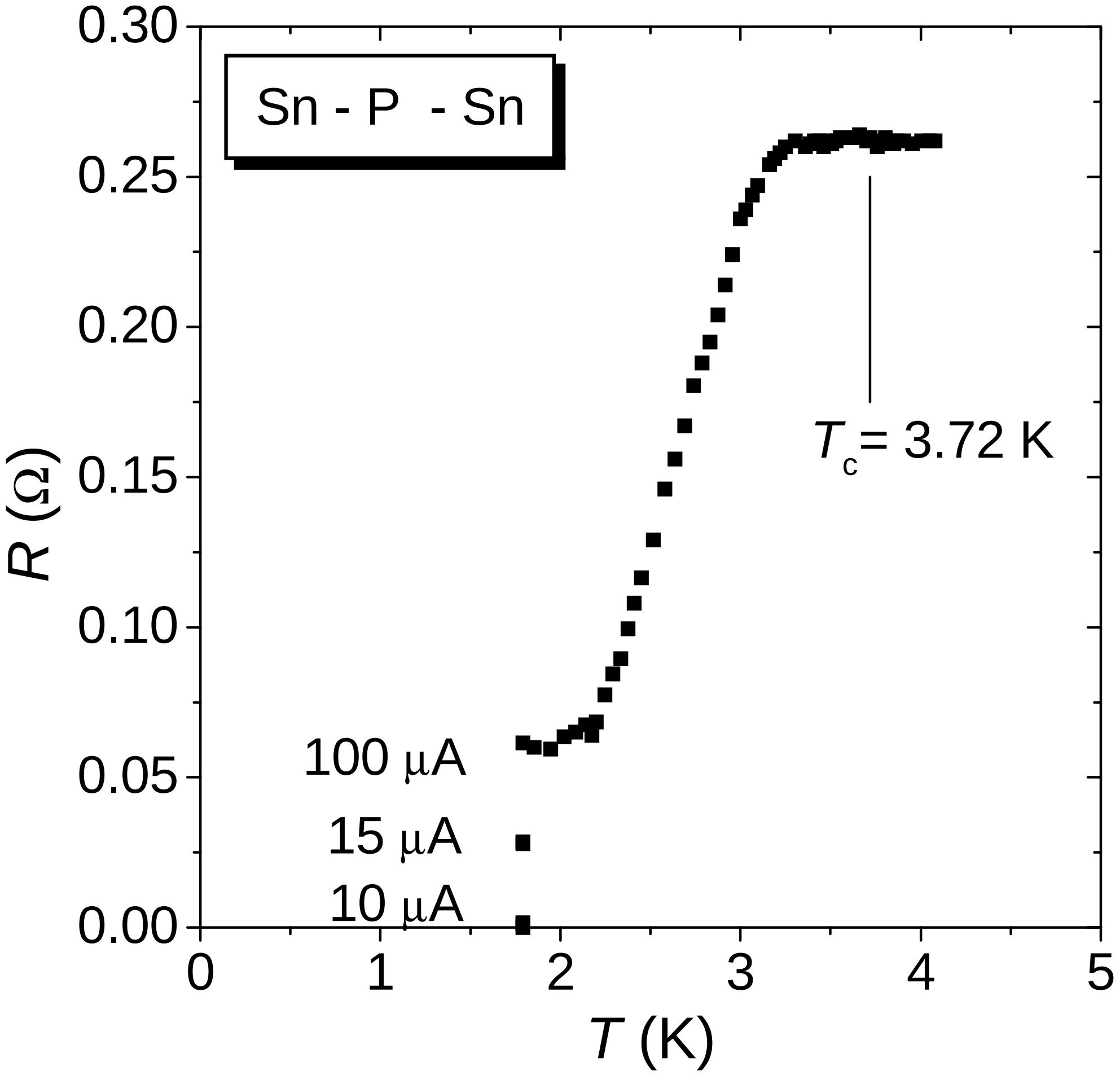

Figure 1a (A. Ionov et al.)

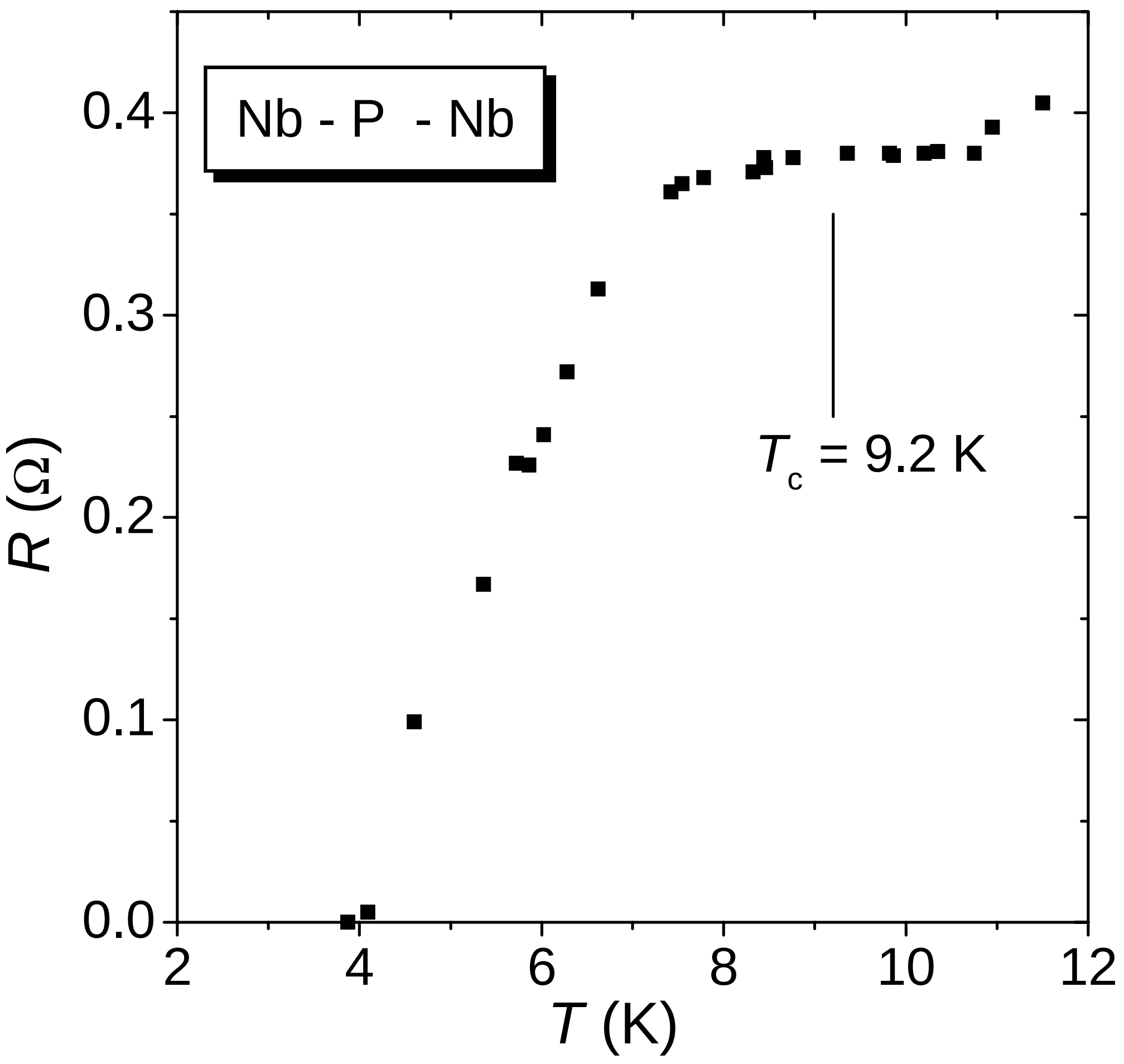

Figure 1b (A. Ionov et al.)

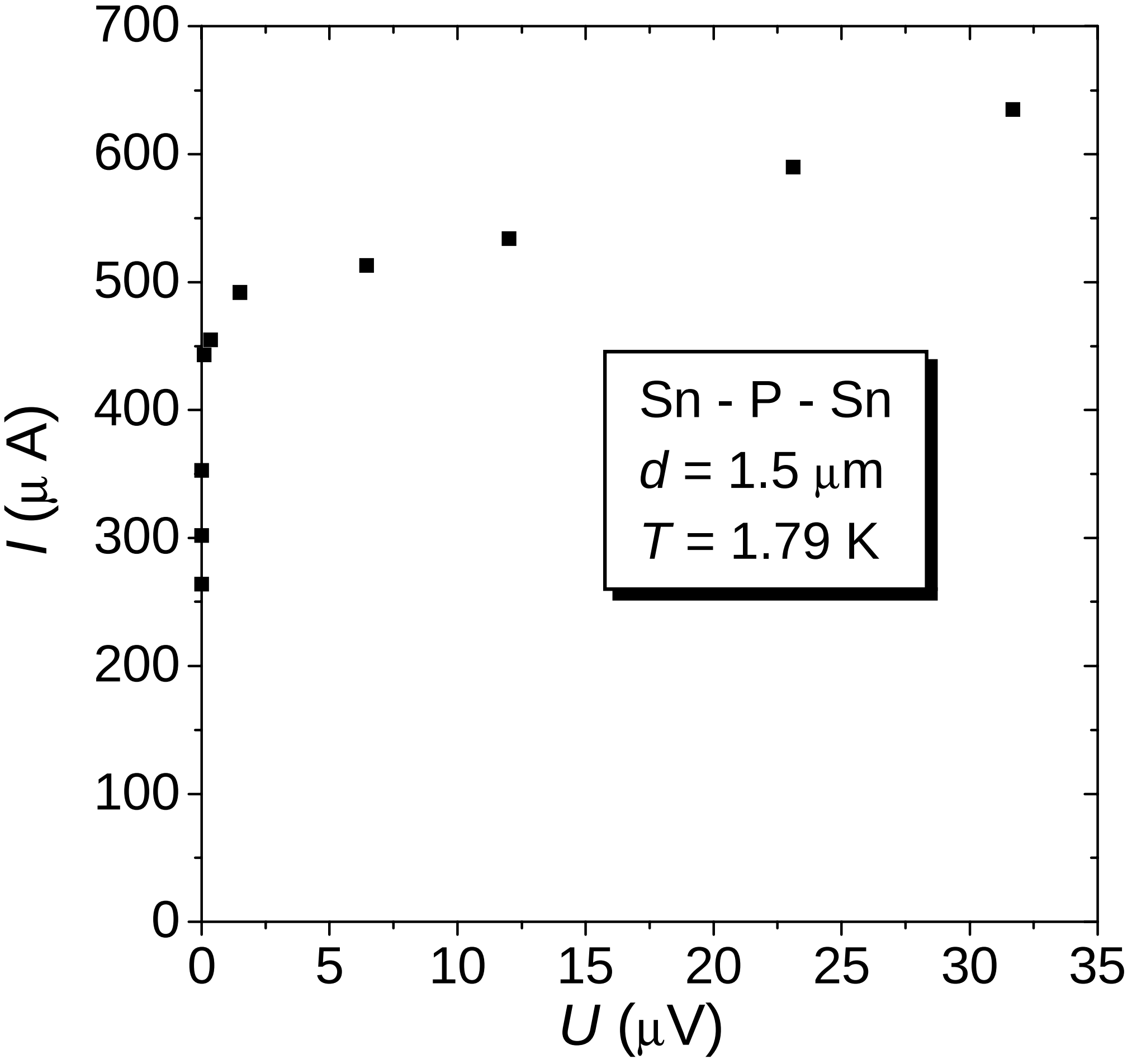

Figure 2 (A. Ionov et al.)

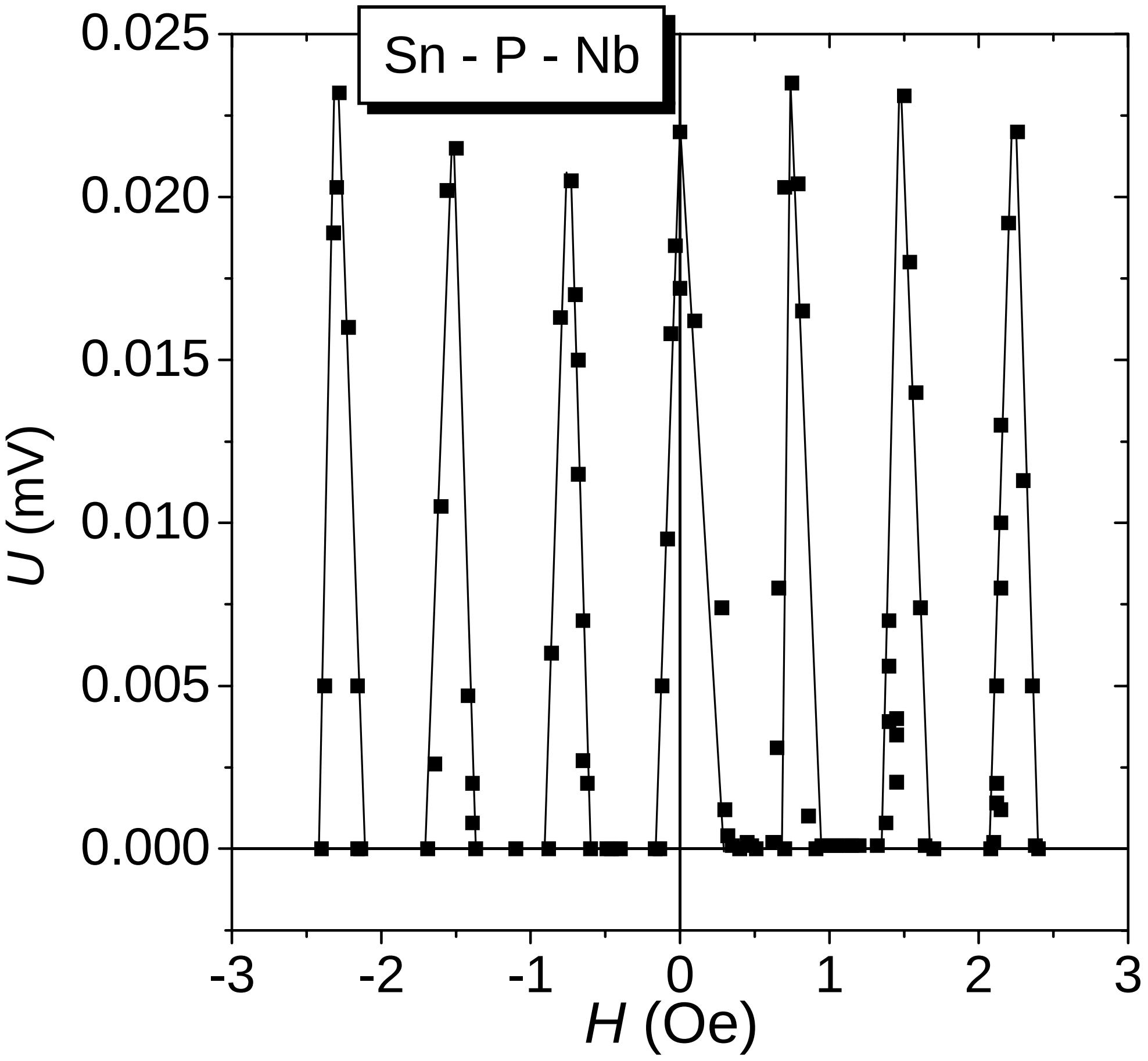

Figure 3 (A. Ionov et al.)